\documentclass{aa}
\usepackage{graphicx}
\usepackage{epsfig}
\usepackage{graphics}
\begin{document}%
   \title{Gamma rays from cloud penetration at the base of AGN jets}

   \author{A.~T. Araudo\inst{1,2,}\thanks{Fellow of CONICET, Argentina},
           V. Bosch-Ramon\inst{3,4}
           \and G.~E. Romero\inst{1,2,}\thanks{Member of CONICET, Argentina}
          }

   \offprints{Anabella T. Araudo: \\ {\em aaraudo@fcaglp.unlp.edu.ar}}
   \titlerunning{Gamma rays from jet-cloud interactions in AGNs}

\authorrunning{Araudo, Bosch-Ramon \& Romero}  \institute{Instituto Argentino de
Radioastronom\'{\i}a, C.C.5, (1894) Villa Elisa, Buenos Aires,
Argentina \and Facultad de Ciencias Astron\'omicas y Geof\'{\i}sicas,
Universidad Nacional de La Plata, Paseo del Bosque, 1900 La Plata,
Argentina \and Max Planck Institut f\"ur Kernphysik, Saupfercheckweg
1, Heidelberg 69117, Germany}

\date{Received / Accepted}


\abstract 
{Dense and cold clouds seem to populate the  broad line region 
surrounding the central 
black hole in AGNs. These clouds could interact with the AGN jet base 
and this could have observational consequences.}
{We want to study the gamma-ray emission produced by these jet-cloud 
interactions, and explore under which conditions 
this radiation would be detectable.}
{We investigate the hydrodynamical properties of jet-cloud 
interactions and the resulting shocks, 
and develop a model to compute the spectral energy
distribution of the emission generated by the particles accelerated in 
these shocks.
We discuss our model in the context of radio-loud AGNs, 
with applications to two representative cases, the low-luminous 
Centaurus A, and the powerful 3C~273.}  
{Some fraction of the jet power can be channelled to gamma-rays, 
which would be likely dominated by synchrotron self-Compton 
radiation, and show typical variability timescales similar to the 
cloud lifetime within the jet, which is longer than several hours. 
Many clouds can interact with the jet simultaneously leading to fluxes 
significantly 
higher than in one interaction, but then variability will be smoothed out.}
{Jet-cloud interactions may produce detectable gamma-rays in non-blazar AGNs,
of transient nature in nearby low-luminous sources like Cen~A, and steady 
in the case of powerful objects of FR~II type.}

\maketitle
%
\section{Introduction}

Active galactic nuclei (AGN) consist of an accreting supermassive
black hole (SMBH) in the center of a galaxy and sometimes present
powerful radio emitting jets (Begelman et al. 1984). Radio-loud AGNs
have continuum emission along the whole electromagnetic spectrum, from
radio to gamma rays (e.g. Boettcher 2007). This radiation basically
comes from the accretion disc and bipolar relativistic jets originated
close to the central SMBH. Radiation of accretion origin can be
produced by the thermal plasma of either an optically-thick
geometrically-thin disc under efficient cooling (Shakura \& Sunyaev
1973), or an optically-thin geometrically-thick corona (e.g. Liang \&
Thompson 1979). The emission from the jets is non-thermal and
generated by a population of relativistic particles likely accelerated
in strong shocks, although other mechanisms are also possible (Rieger
et al. 2007). This non-thermal emission is thought to be produced
through synchrotron and inverse Compton (IC) processes
(e.g. Ghisellini et al. 1985), although hadronic models have been also
considered to explain gamma-ray detections (e.g. Mannheim 1993,
M\"ucke \& Protheroe 2001, Aharonian 2002). 

In addition to continuum radiation, AGNs also present optical and
ultra-violet lines. Some of these lines are broad, emitted by gas
moving with velocities $v_{\rm g}>1000$~km~s$^{-1}$ and located in a
small region close to the SMBH, the so-called broad line
region (BLR). The structure of this region is not well known but some
models assume that the material in the BLR could be formed by dense
clouds confined by a hot ($T \sim 10^8$~K) external medium (Krolik et
al. 1981) or by magnetic fields (Rees 1987). These clouds would be
ionized by photons from the accretion disc producing the observed
emission lines, which are broad because of the cloud motion within the
SMBH potential well. An alternative model proposes that the broad
lines are produced in the chromosphere of evolved stars (Penston 1988)
present in the nuclear region of AGNs.  

The presence of material surrounding the base of the jets in AGNs
makes jet-medium interactions likely. For instance, the interaction of
BLR clouds with a jet in AGNs was already suggested by Blandford \&
K\"onigl (1979) as a mechanism for knot formation in the radio galaxy
M87. Also, the gamma-ray production due to the interaction of a cloud
from the BLR with a proton beam or a massive star
with a jet were studied in the context of AGNs by Dar \& Laor (1997) 
and Bednarek \& Protheroe (1997), respectively. 

In this work, we study the interaction of BLR clouds with the innermost
jet in an AGN and its observable consequences at high
energies. The approach adopted is similar to that followed in Araudo
et al. (2009) for high-mass microquasars (for a general comparison
between these sources and AGNs see Bosch-Ramon 2008),
where the interaction of stellar wind clumps of the companion star 
with the microquasar jet was studied.  Under magnetic fields below
equipartition with the jet kinetic energy (i.e. the jet should be
matter dominated), cloud penetration will lead to the formation of a
relativistic bow shock in the jet and a slow shock inside the
cloud. Electrons and protons can be efficiently accelerated in the bow
shock and produce non-thermal emission, in situ via synchrotron and
synchrotron self-Compton (SSC) mechanism, and in the cloud through 
proton-proton ($pp$)
collisions. For magnetic fields well below equipartition, the SSC
component becomes the dominant electron cooling channel, that leads
to significant gamma-ray production. Since the bow shock downstream is
almost at rest in the laboratory reference frame (RF), this emission
will not be significantly boosted. The resulting spectrum and the
achieved luminosities in one jet-cloud interaction depend strongly on
the magnetic field, the location of the interaction region, 
the cloud size, and the
jet luminosity. However, many clouds could be inside the jet
simultaneously, and then the BLR global properties, like size and
total number of clouds, would also be relevant. Depending on whether
one cloud or many of them penetrate into the jet, the lightcurve will
be flare-like or rather steady, respectively.

In order to explore the radiative outcomes of jet-cloud interactions
in AGNs, we apply our model to both Faranoff-Riley galaxies I (FR~I)
and II (FR~II). In particular, we consider Centaurus A (Cen A) and 
3C~273, the nearest FR~I
and a close and very bright flat spectrum radio quasar (with
FR~II as parent population), as illustrative cases. Although in FR~I
the BLR is not well-detected, clouds with similar characteristics to
those found in FR~II galaxies may surround the SMBH (Wang et
al. 1986, Risaliti 2009). Cen~A has been detected at high- (HE)
(Hartman et al. 1999; Abdo et al. 2010) and very high-energy (VHE) gamma
rays (Aharonian et al. 2009), whereas 3C~273 has been
detected so far only at HE gamma rays (Hartman et al. 1999;
Abdo et al. 2010). We have computed the contribution of jet-cloud
interactions to the gamma-ray emission in these sources, and estimated
the gamma-ray luminosity in a wide range of cases. We find that
gamma rays from jet-cloud interactions could be detectable by
present and future instrumentation in nearby low-luminous AGNs at HE
and VHE, and for powerful and nearby quasars only at HE, since the VHE
radiation is absorbed by the dense nuclear photon fields. In the case
of sources showing boosted gamma rays (blazars), the isotropic
radiation from jet-cloud interactions will be masked by the jet beamed
emission, which will not be the case in non-blazar sources.

The paper is organized as follows: in Sect.~\ref{jet-cloud}, the
dynamics of jet-cloud interactions is described;
in Sect.~\ref{acc}, a model for particle acceleration and emission is
presented for one interaction, whereas 
in Sect.~\ref{Many-clouds} the case of many clouds interacting with the
jet is considered; 
in Sects.~\ref{FRI} and \ref{FRII}, the
model is applied to FR~I and FR~II galaxies, focusing on the
sources Cen A and 3C~273; finally, in  Sect.~\ref{disc}, the
results of this work are summarized and discussed. We adopt cgs units
through-out the paper.   

\section{The jet-cloud interaction}\label{jet-cloud}

\begin{figure}
\begin{center}
\includegraphics[angle=0, width=0.35\textwidth]{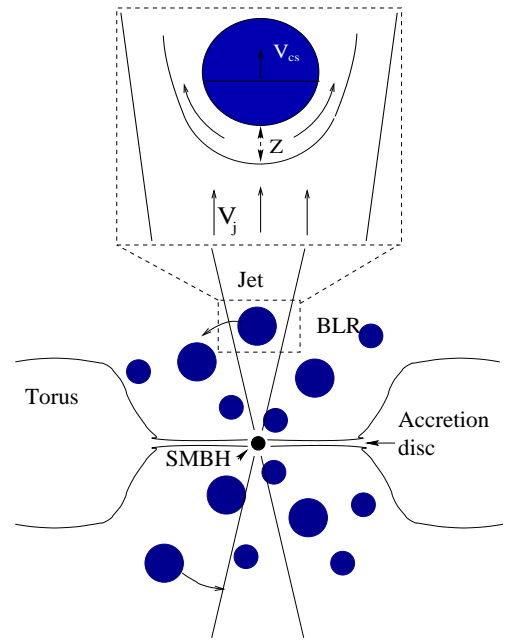}
\caption{Sketch, not to scale, of an AGN  at the spatial
scales of the BLR region. In the top part of the figure, 
the interaction between a cloud and the jet is also shown.}
\label{blr}
\end{center}
\end{figure}

Under certain combinations of the jet ram pressure, and the cloud
size and density, cloud-jet penetration is expected to occur. The
details of the penetration process itself are complex. Here we do
not treat them in detail, but just assume that penetration occurs if
certain conditions are fulfilled. For low magnetic fields, 
 a cloud inside the jet may represent  a
hydrodynamic situation in which a supersonic flow interacts
with a body of approximately
spherical shape at rest. The cloud, as long as it has not been accelerated by
the jet ram pressure up to the jet speed ($v_{\rm j}$), produces a
perturbation in the jet medium in the form of a steady bow shock
roughly at rest in the laboratoty RF, with a velocity with respect to
the jet RF approximately equal to $v_{\rm j}$. Since the cloud is not
rigid, a wave propagates also through it.  Since the cloud temperature
is much lower, and the density much higher than in the jet, this wave
will be still supersonic but much slower than the bow shock. The jet
pressure exerts a force on the cloud leading to cloud 
acceleration along the axis, hydrodynamical instabilites and,
eventually, cloud fragmentation. In the following, the jet-cloud
interaction is described. Further discussion, and a proper account of
the literature, can be found in Araudo et al. (2009). Sketchs of the
jet-cloud interaction and the jet-BLR scenario are shown in
Fig.~\ref{blr}.

\begin{table}[]
\begin{center}
\caption{Values assumed in this work for BLR clouds and jets.}
\label{const}
\begin{tabular}{ll} 
\hline  Description   & Value \\  
\hline Cloud size  & $R_{\rm c} = 10^{13}$~cm \\ 
Cloud density  & $n_{\rm c} = 10^{10}$~cm$^{-3}$ \\ 
Cloud velocity  & $v_{\rm c} = 10^{9}$~cm~s$^{-1}$ \\ 
Cloud temperature  & $T_{\rm c} = 2\times10^{4}$~K \\ 
Jet Lorentz factor & $\Gamma = 10$ \\ 
Jet half-opening angle & $\phi\approx 6^{\circ}$ \\ 
\hline
\end{tabular}
\end{center}
\end{table}

We adopt here clouds with typical density $n_{\rm
c}=10^{10}$~cm$^{-3}$ and size $R_{\rm c}=10^{13}$~cm (Risaliti 2009).
 The velocity of the cloud is taken
$v_{\rm c}=10^9$~cm~s$^{-1}$ (Peterson 2006). The jet Lorentz
factor is fixed to $\Gamma=10$, implying $v_{\rm j}\approx c$, with a
half-opening angle $\phi\approx 6^{\circ}$, i.e. the jet radius/height
relation fixed to $R_{\rm j}=\tan(\phi)\,z=0.1\,z$. All these
parameters are summarized in Table~\ref{const} and
will not change along the paper;
from them, the jet density $n_{\rm j}$ in the
laboratoty RF can be estimated:
\begin{eqnarray} 
n_{\rm j} & = &  \frac{L_{\rm j}}{(\Gamma-1)\,m_{\rm p}\,c^3 \sigma_{\rm j}} 
\approx 8\times 10^{4}\left(\frac{L_{\rm j}}{10^{44}\,\rm{erg\, s^{-1}}}\right)
\nonumber\\ 
{} & \times &  
\left(\frac{\Gamma-1}{9}\right)^{-1}
\left(\frac{z}{10^{16}\,{\rm cm}}\right)^{-2} \,\rm{cm^{-3}},
\end{eqnarray}
where $\sigma_{\rm j} = \pi R_{\rm j}^2$ and 
$L_{\rm j}$ is the kinetic power of the matter dominated jet. 

The jet ram pressure should not destroy the cloud before it 
has fully entered into the jet. This means that
the time required by the cloud to penetrate into the jet should be:
\begin{equation} 
t_{\rm c}\sim \frac{2 R_{\rm c}}{v_{\rm c}} =
2\times 10^4\left(\frac{R_{\rm c}}{10^{13}\,\rm{cm}}\right)
\left(\frac{v_{\rm c}}{10^9\,\rm{cm\, s^{-1}}}\right)^{-1}\,{\rm s}\,,
\end{equation} 
should be shorter than the cloud lifetime inside the jet. To estimate 
this cloud lifetime, let us
compute first the time required by the shock in the cloud to cross it 
($t_{\rm cs}$).  The
velocity of this shock, $v_{\rm cs}$, can be derived making equal the 
jet and the cloud shock ram pressures: 
$(\Gamma-1)\,n_{\rm j}\,m_{\rm p}\,c^2=n_{\rm c}\,m_{\rm p}\,v_{\rm cs}^2$,
valid as long as $v_{\rm cs}\ll c\,$. Then:
\begin{eqnarray} 
v_{\rm cs} & \sim & \chi^{-1/2}\, c\sim
3\times10^8\left(\frac{n_{\rm c}}{10^{10}\,\rm{cm^{-3}}}\right)^{-1/2}
\nonumber\\ 
{} & \times &  \left(\frac{z}{10^{16}\,{\rm
cm}}\right)^{-1}\left(\frac{L_{\rm j}}{10^{44}\,\rm{erg\,
s^{-1}}}\right)^{1/2} \, \rm{cm\,s^{-1}}, 
\end{eqnarray} 
where $\chi$ is the cloud to jet density ratio, 
$n_{\rm c}/n_{\rm j}(\Gamma-1)$.  This yields a clould shocking time:
\begin{eqnarray} 
t_{\rm cs} &\sim& \frac{2R_{\rm c}}{v_{\rm cs}}\simeq
7\times10^4 \left(\frac{R_{\rm c}}{10^{13}\, \rm{cm}} \right)\,
\left(\frac{n_{\rm c}}{10^{10}\,\rm{cm^{-3}}}\right)^{1/2} \nonumber\\
{}& \times {}& \left(\frac{z}{10^{16}\,{\rm
cm}}\right)\left(\frac{L_{\rm j}}{10^{44}\,\rm{erg\,
s^{-1}}}\right)^{-1/2}\,{\rm s}\,.
\end{eqnarray} 
Therefore, for a  penetration time ($t_{\rm c}$) at least as short 
as $\sim t_{\rm cs}$, the
cloud will remain being an effective obstacle for the jet
flow. Setting $t_{\rm c}\sim t_{\rm cs}$, we obtain a minimum value
for $\chi$ and hence for $z$. 

Hydrodynamical instabilities produced by the interaction with the jet 
material will affect the cloud. First of all, the jet exerts a
force on the cloud through the contact discontinuity. The acceleration
applied to the cloud can be estimated from the jet ram pressure
$P_{\rm j}$, and the cloud section $\sigma_{\rm c}\sim \pi\,R_{\rm c}^2$
and mass $M_{\rm c}\sim (4/3)\,\,\pi R_{\rm c}^3\,n_{\rm j}\,m_{\rm p}$:
\begin{equation} 
g=\frac{P_{\rm j}\,\sigma_{\rm c}}{M_{\rm c}}\sim
\frac{3}{4}\frac{c^2}{\chi\,R_{\rm c}}=\frac{3}{2}\,
\frac{v_{\rm cs}}{t_{\rm cs}}\,. 
\end{equation}
Given the acceleration exerted by the jet in the cloud,  
Rayleigh-Taylor (RT) instabilities will develop in the cloud, at the
jet contact discontinuity, with timescale:
\begin{equation} 
t_{\rm RT} \sim \sqrt{\frac{l}{g}}=
\sqrt{\frac{4\,\chi\,l\,R_{\rm c}}{3\,c^2}}\,, 
\end{equation} 
where the instability length $l$ is the spatial scale of the
perturbation. For perturbations of the size of the cloud: $l\sim
R_{\rm c}$, which are those associated to cloud significant
disruption, one gets $t_{\rm RT}\sim t_{\rm cs}$.

In addition to RT instabilities, Kelvin-Helmholtz (KH) instabilities
also grow in the cloud walls in contact with  shocked jet material
that surrounds the cloud. Accounting for the high relative
velocity, $v_{\rm rel}\la v_{\rm j}$, one obtains:
\begin{equation} 
t_{\rm KH}\ga \sqrt{\frac{l}{g_{\rm rel}}}=\frac{\chi\,l}{c}\,, 
\end{equation} 
where $g_{\rm rel}\sim c^2/\chi\,l$.  For $l\sim R_{\rm c}$, we obtain
$t_{\rm KH}\ga t_{\rm cs}$.  In the previous estimates of $t_{\rm
RT}$ and $t_{\rm KH}$ we have not taken into account the effect of the
magnetic field (e.g. Blake 1972), since we assume that it is
dynamically negligible. We note that, given $g$, the time to
accelerate the cloud up to the shock velocity $v_{\rm cs}$ is $\sim
t_{\rm cs}$.  However, the timescale to accelerate the cloud up to
$v_{\rm j}$ is $\gg t_{\rm cs}$ provided that $v_{\rm j}\gg v_{\rm
cs}$, and before, the cloud will likely fragment. 

Finally, there are two additional timescales also relevant for our
study, the bow-shock formation time, $t_{\rm bs}$, and the time
required by the cloud to cross the jet, $t_{\rm j}$. The timescale
$t_{\rm bs}$ can be roughly estimated assuming that  the shock
downstream has a cylindrical shape with one of the bases being the bow
shock, relativistic shock jump conditions, equal particle injection
and escape rates,  and a escape velocity similar to the sound speed
$\sim c/\sqrt{3}$ (for a  relativistic plasma). This yields a shock-cloud
separation distance of $Z \sim 0.3\,R_{\rm c}$, which implies: 
\begin{equation} 
t_{\rm bs} \sim \frac{Z}{c}=10^2
\left(\frac{R_{\rm c}}{10^{13}\, \rm{cm}} \right)\,{\rm s}\,.
\end{equation} 
Since in general $t_{\rm bs}\ll t_{\rm cs}$, we can assume that the bow shock 
is in the steady regime. The jet crossing time 
$t_{\rm j}$ can be characterized by:
\begin{equation} t_{\rm j}\sim \frac{2 R_{\rm j}}{v_{\rm c}} =
2\times10^6 \left(\frac{z}{10^{16}\,{\rm cm}}\right)
\left(\frac{v_{\rm c}}{10^9\,\rm{cm\, s^{-1}}}\right)^{-1} \,{\rm s}\,.
\end{equation}  
Note that if the cloud lifetime is $<t_{\rm j}$, the number of 
clouds inside the jet will be smaller
than expected just from the BLR properties. 

\begin{figure}
\includegraphics[angle=270, width=0.5\textwidth]{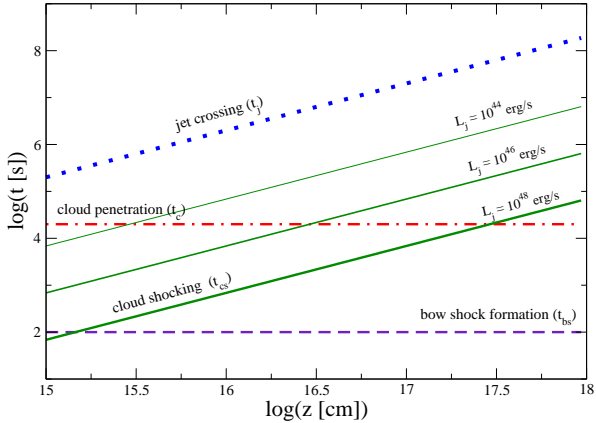}
\caption{The
jet crossing (blue dotted line), cloud penetration (red dot-dashed line),
bow-shock formation (violet dashed  line) and cloud shocking 
(green solid lines) times are plotted; all of them have been calculated using 
the values given in Table~\ref{const}. The time $t_{\rm cs}$ is plotted for 
$L_{\rm j}=10^{44}$, $10^{46}$ and $10^{48}$~erg~s$^{-1}$.}
\label{timescales}
\end{figure}

In order to summarize the discussion of the dynamics of the jet-clump
interaction, we plot in Fig.~\ref{timescales} the $t_{\rm cs}$ (for
different $L_{\rm j}$), $t_{\rm j}$, $t_{\rm c}$, and $t_{\rm bs}$ as a
function of $z$. As shown in the figure, for some values of $z$ and
$L_j$ the cloud could be destroyed by the jet before full penetration,
i.e. $t_{\rm cs}<t_{\rm c}$. This is a constraint to determine the
height $z$ of the jet at which the cloud can penetrate into it. Note
also that, in general, $t_{\rm bs}$ is much shorter than any other
timescale.

\subsection{The interaction height}
\label{InteractionHeight}

The cloud can fully penetrate into the jet if the cloud lifetime after jet
impact is longer than the penetration time (the weaker condition that
jet lateral pressure is $<n_{\rm c}\,m_{\rm p}\,v_{\rm c}^2$ is then
automatically satisfied). This determines the minimum interaction
height, $z_{\rm int}$, to avoid cloud disruption before full
penetration. Also, this interaction cannot occur below 
the jet formation region,
$z_{\rm 0}\sim 100\,R_{\rm g}\approx 1.5\times 10^{15}\,(M_{\rm
bh}/10^8\,M_{\odot})$~cm (Junor et al. 1999). For BLR-jet interaction
and cloud penetration to occur, the size of the BLR should be 
$R_{\rm blr}>z_0$ and $z_{\rm int}$. 

The lifetime of the cloud depends on the fragmentation time, which is
strongly linked to, but longer than, $t_{\rm cs}$. 
The value of $z_{\rm int}$ can be
estimated then setting $t_{\rm c}\la t_{\rm cs}$, since the cloud
should enter the jet before being significantly distorted by the
impact of the latter. Once shocked, the cloud can suffer lateral
expansion and conduction heating, which can speed up fragmentation due
to instabilities. In this work, we choose
$z_{\rm int}$ fixing $t_{\rm cs} = 2\,t_{\rm c}$:
\begin{eqnarray}
z_{\rm int}  & \approx & 5\times10^{15}
\left(\frac{v_{\rm c}}{10^9\,\rm{cm\,s^{-1}}}\right)^{-1}
\left(\frac{n_{\rm c}}{10^{10}\,\rm{cm^{-3}}}\right)^{-1/2} \nonumber\\
{} & {\times} &
\left(\frac{L_{\rm j}}{10^{44}\,\rm{erg\, s^{-1}}}\right)^{1/2} \,\rm{cm}.
\end{eqnarray}  
We note that the available power in the bow shock is 
$L_{\rm bs}\sim (\sigma_{\rm c}/\sigma_{\rm j})\,L_{\rm j}\propto z^{-2}$. 
Therefore, the most luminous individual jet-cloud 
interaction  will take place at $z\sim z_{\rm int}$.

The BLR size can be estimated through 
an empirical relation obtained from sources with a well stablished BLR,
i.e. FR~II radio galaxies.
This relation is in general of the type 
$R_{\rm blr} \propto L_{\rm blr}^{\alpha}$, where $L_{\rm blr}$ is the 
luminosity of the BLR and $\alpha \sim 0.5 - 0.7$ (e.g. Kaspi et al. 2005,
2007; Peterson et al. 2005; Bentz et al. 2006). 
In this paper we  use the following relations: 
\begin{equation}
\label{Rblr}
R_{\rm blr}\sim 6\times 10^{16} 
\left(\frac{L_{\rm blr}}{10^{44}\,\rm{erg\,s^{-1}}}\right)^{0.7}\,\rm{cm},
\end{equation}
and
\begin{equation}
\label{Rblr_07}
R_{\rm blr}\sim 2.5\times 10^{16} 
\left(\frac{L_{\rm blr}}{10^{44}\,\rm{erg\,s^{-1}}}\right)^{0.55}\,\rm{cm},
\end{equation}
from Kaspi et al. (2005, 2007).
%

\begin{figure}
\includegraphics[angle=270, width=0.5\textwidth]{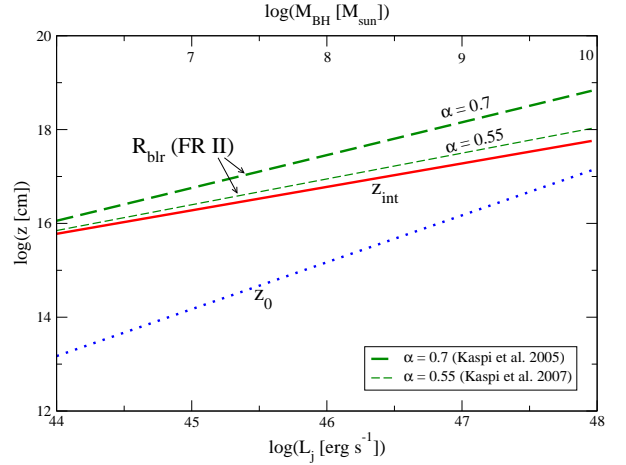}
\caption{The interaction height, $z_{\rm int}$ (red solid line), and the size 
of the BLR, $R_{\rm blr}$ (green dashed line), for different values 
of $L_{\rm j}$ (bottom horizontal axis). We have derived 
$R_{\rm blr}(L_{\rm j})$ fixing $L_{\rm blr} = 0.1\,L_{\rm j}$ and 
plotted $R_{\rm blr}$ using Eqs.~(\ref{Rblr}) and (\ref{Rblr_07}).
In the same plot, the height of the jet base, $z_0$ (blue dotted line), 
is plotted as a function of $M_{\rm BH}$ (top horizontal axis).}
\label{zint-rblr}
\end{figure}

In Fig.~\ref{zint-rblr} we show the relation of $z_{\rm int}$ and
$R_{\rm blr}$ with $L_{\rm j}$, assuming that $L_{\rm blr}$ is a
10\% of the disc luminosity, which is taken here equal to $L_{\rm j}$. As
seen in the figure, for reasonable parameters, the condition $z_{\rm
int}<R_{\rm blr}$ is fulfilled for a wide range of $L_{\rm j}$.
Figure~\ref{zint-rblr} also shows the relation between $z_0$ and
$M_{\rm bh}$, which shows that for $M_{\rm bh}\ga 10^9$~$M_{\odot}$
the jet could be even not (fully) formed at the BLR scales at the
lowest $L_{\rm j}$-values. 

\section{Non-thermal particles and their emission}\label{acc}

In the bow and cloud shocks, particles can be accelerated through
diffusive shock acceleration (Bell 1978). However, the bow shock
should be more efficient accelerating particles 
than the shock in the cloud because 
$v_{\rm bs}\gg v_{\rm cs}$. In addition, the cloud shock luminosity is smaller
than the bow-shock luminosity by $\sim 1/(2\chi^{1/2})$. For these
reasons, we focus here on the particle acceleration in the bow shock. 
In this section, we briefly
describe the injection and evolution of particles, and their emission,
remarking those aspects that are specific to AGNs. The details of the
emitting processes considered here (synchrotron, IC and $pp$
interactions) can be found in Araudo et al. (2009) and references
therein. 

First, one can estimate the non-thermal luminosity, $L_{\rm nt}$,
injected at $z_{\rm int}$ in the bow shock in the form of relativistic
electrons or protons:
\begin{eqnarray}
L_{\rm nt}& = &\eta_{\rm nt} L_{\rm bs} \approx 4\times10^{39} 
\left(\frac{\eta_{\rm nt}}{0.1}\right)
\left(\frac{R_{\rm c}}{10^{13}\rm{cm}}\right)^2\\\nonumber
{}&\times& 
\left(\frac{L_{\rm j}}{10^{44}\,\rm{erg\,s^{-1}}}\right)\,\rm{erg\,s^{-1}}.
\end{eqnarray}
Then, the accelerator/emitter magnetic field in the bow-shock RF ($B$) 
can be determined relating 
$U_{\rm B} = \eta_{\rm B} U_{\rm nt}$,
where $U_{\rm B}=B^2/8\pi$ and $U_{\rm nt}=L_{\rm nt}/(\sigma_{\rm c}c)$ 
are the magnetic and the non-thermal energy densities,
respectively. For leptonic emission and to avoid supression of the IC channel, 
high gamma-ray outputs require $\eta_{\rm B}$ well below 1. In this 
context, $B$ can be parametrized as follows:
\begin{equation}
B\approx 10 \left(\frac{\eta_{\rm B}}{0.01}\right)
\left(\frac{v_{\rm c}}{10^{9}\, \rm{cm\,s^{-1}}}\right)^2
\left(\frac{n_{\rm c}}{10^{10}\, \rm{cm^{-3}}}\right)\,{\rm G}\,.
\end{equation}
Regarding the acceleration mechanism, since the bow shock is 
relativistic and the treatment of such a shocks is complex 
(see Achterberg et al. 2001), we adopt the following prescription 
for the acceleration rate: 
\begin{equation}
\dot{E}_{\rm acc}=0.1\,q\,B\,c\,, 
\end{equation}
similar to that in the relativistic
termination shock of the Crab pulsar wind (de Jager et al. 1996). 

Particles suffer different losses that balance the energy gain from
acceleration. The electron loss mechanisms are escape downstream,
relativistic Bremsstrahlung, synchrotron emission, and external
Compton (EC) and SSC.  We note that $B$, $L_{\rm nt}$  and the 
accelerator/emitter size at $z_{\rm int}$ are constant for different
$L_{\rm j}$ and fixed $\eta_{\rm B}$ and $\eta_{\rm nt}$, and only
$L_{\rm blr}$ and $L_{\rm d}$ are expected to change with $L_{\rm j}$. 
Therefore, as
long as the external photon fields are negligible, the maximum
electron energy at $z_{\rm int}$ does not change for different jet 
powers.

\begin{figure}
\includegraphics[angle=270, width=0.5\textwidth]{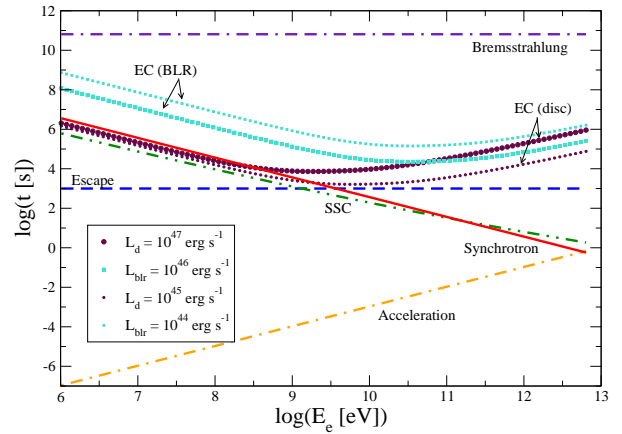}
\caption{Acceleration gain (orange dot-dashed line), escape (blue dashed line)
and cooling lepton timescales are plotted. 
SSC (green two-dot-dashed line) is plotted when the steady state is reached
and EC for the both BLR (turquoise square line) and disc (maroon dotted line) 
photon fields are shown for the 
conditions of faint (BLR: $10^{44}$; disc: $10^{45}$~erg~s$^{-1}$) and 
bright sources (BLR: $10^{46}$; disc: $10^{47}$~erg~s$^{-1}$). 
Synchrotron (red solid line) and relativistic Bremsstrahlung (violet 
dot-two-dashed line) are also plotted. We have fixed here 
$\eta_{\rm B} = 0.01$. For protons, the 
$pp$ cooling timescale is not shown for clarity, although it would 
be similar to that of relativistic bremsstrahlung.}
\label{losses}
\end{figure}

In Fig.~\ref{losses}, the leptonic cooling timescales are plotted together 
with the escape time and the acceleration time for a bow
shock located at $z_{\rm int}$. A value for $\eta_{\rm B}$ equal to 
0.01 has been adopted. The SSC cooling timescale is plotted 
for the steady state. The escape time downstream the
relativistic bow shock is taken as:
\begin{equation} 
t_{\rm esc}\sim \frac{3\,R_{\rm c}}{c} = 10^3\,
\left(\frac{R_{\rm c}}{10^{13}\, \rm{cm}}\right)\,{\rm s}\,. 
\end{equation}
Synchrotron and EC/SSC are the dominant processes in the high-energy
part of the electron population, relativistic bremsstrahlung is
negligible for any energy, and electron escape is relevant in the
low-energy part. This yields a break in the electron energy
distribution at the energy at which the synchrotron/IC time and the
escape time are equal. The Thomson to KN transition is clearly seen in the
EC cooling curves, but is much smoother in the SSC case. The 
maximun electrons energy are around several TeV. Given the
similar cooling timescale for electrons via relativistic
bremsstrahlung and protons through $pp$ collisions ($t_{\rm
brems/pp}\sim 10^{15}/n$~s, where $n$ is the target density), protons
will not cool efficiently in the bow shock. Photomeson production can
also be discarded as a relevant proton cooling mechanism in the bow
shock due to the relatively low achievable proton energies and photon
densities.
 The maximum proton energy
of protons is constrained making equal the acceleration time and the
time needed to diffuse out of the bow shock. Assuming Bohm diffusion,
$t_{\rm diff}=3\,Z^2/2\,r_{\rm g}\,c$ (where $r_{\rm g}$ is the
particle gyroradius), maximum proton energy is:
\begin{equation}
E_{\rm p}^{\rm max} \sim 0.1\,q\,B\,R_{\rm c}=
5\times 10^3\,\left(\frac{B}{10\,{\rm G}}\right)\,\left(\frac{R_{\rm
c}}{10^{13}\,{\rm cm}}\right)\,{\rm TeV}\,.  
\end{equation}

Electrons are injected in the bow shock region following a power-law
in energy of index 2.2 (Achterberg et al. 2001) with an exponential
cutoff at the maximum electron energy.  The injection luminosity is
$L_{\rm nt}$. To first order, the electron evolution can be
computed assuming homogeneous conditions, following therefore a
one-zone approximation with all the mentioned cooling and escape
processes.  The formulae for all the relevant radiative mechanisms, as well as
the solved electron evolution differential equation, can be found in
Araudo et al. (2009). In some cases, SSC is the dominant cooling
channel at high energies. In that case, the calculations have to be 
done numerically
splitting the evolution of the electron population in different time
steps. In each step, the radiation field is updated accounting for the
synchrotron emission produced in the previous step until the steady
state is reached. The duration of each step should  be shorter for the
earlier phases of the evolutionary state to account properly for the
rise of the synchrotron energy density in the emitter.

Once the steady-state electron distribution in the bow shock is
computed, the spectral energy distribution (SED) of the non-thermal
radiation can be calculated. The synchrotron self-absorption effect
has to be taken into account, which will affect the low energy band of
the synchrotron emission. At gamma-ray energies, photon-photon
absorption due to the disc and the BLR radiation are to be considered,
the internal absorption due to synchrotron radiation being negligible. 
Given the typical
BLR and disc photon energies, $\sim 10$~eV and $\sim 1$~keV,
respectively, gamma rays beyond 1~GeV and 100~GeV can be strongly
affected by photon-photon absorption. On the other hand, for most of 
cases photons between 100~MeV
and 1~GeV energies will escape the dense disc photon field.

Although proton cooling is negligible in the bow-shock region, it may
be significant in the cloud. Protons can penetrate into the cloud if
$t_{\rm esc}>t_{\rm diff}$, which yields a minimum energy to reach the cloud of
$E_p\sim 0.4\,E_p^{\rm max}$. These protons will radiate in the form of
gamma rays only a fraction $\sim 3\times 10^{-4}\,(R_{\rm
c}/10^{13}\,{\rm cm})(t_{\rm pp}/10^5\,{\rm s})^{-1}$  of their
energy, wich makes the process rather unefficient. The reason is that
these protons cannot be efficiently confined and cross the cloud
at a velocity $\sim c$. For further details of the proton energy distribution
in the cloud, see Araudo et al. (2009).

\section{Many clouds inside the jet}\label{Many-clouds}

Clouds fill the BLR, and many of them can simultaneously
be inside the jet at different $z$, each of them producing non-thermal
radiation. Therefore, the total luminosity can be much larger than the
one produced by just one interaction, which is $\sim L_{\rm nt}$.  The
number of clouds within the jets, at $z\le R_{\rm blr}$, can be
computed from the jet ($V_{\rm j}$) and
cloud ($V_{\rm c}$) volumes, resulting in:
\begin{equation}
\label{N_clouds}
N_{\rm c}^{\rm j} = 2\, f\, \frac{V_{\rm j}}{V_{\rm c}}\sim
9\left(\frac{L_{\rm j}}{10^{44}\,\rm{erg s^{-1}}}\right)^{2}
\left(\frac{R_{\rm c}}{10^{13}\,\rm{cm}}\right)^{-3},
\end{equation}
where the factor 2 accounts for the two jets and 
$f \sim 10^{-6}$ is the filling factor of clouds in the whole
BLR (Dietrich et al. 1999). Actually, $N_{\rm c}^{\rm j}$ 
is correct if one  neglects
that the cloud disrupts and fragments, and eventually dilutes inside
the jet. For instance, Klein et al. (1994) estimated a shocked cloud
lifetime in several $t_{\rm cs}$, and Shin et al. (2008) found that
even a weak magnetic field in the cloud can significantly increase its
lifetime. Finally, even under cloud fragmentation, strong bow shocks
can form around the cloud fragments before these have accelerated
close to $v_{\rm j}$. All this makes the real number of interacting
clouds inside the jet hard to estimate, but it should be between
$(t_{\rm cs}/t_{\rm j})\,N_{\rm c}^{\rm j}$ and $N_{\rm c}^{\rm j}$. 

The presence of many clouds inside the jet, not only at $z_{\rm int}$
but also at higher $z$, implies that the total non-thermal luminosity
available in the BLR-jet intersection region is:
\begin{eqnarray}
\label{Lrad_tot}
L_{\rm nt}^{\rm tot} &\sim &2\, \int^{R_{\rm blr}} 
\frac{{\rm d}N_{\rm c}^{\rm j}}{{\rm d}z} L_{\rm nt}(z)\, {\rm d}z\nonumber \\ 
{}&\sim & 2\times10^{40} \left(\frac{\eta_{\rm nt}}{0.1}\right) 
\left(\frac{R_{\rm c}}{10^{13}\,\rm{cm}}\right)^{-1}
\left(\frac{L_{\rm j}}{10^{44}\,\rm{erg\,s^{-1}}}\right)^{1.7}
\end{eqnarray} 
where ${\rm d}N_{\rm c}^{\rm j}/{\rm d} z$ is the number of clouds
located in a jet volume  ${\rm d}V_{\rm j}=\pi\,(0.1z)^2\,{\rm
d}z$. In both Eqs.~\ref{N_clouds} and \ref{Lrad_tot}, 
$L_{\rm blr}$
has been fixed to $0.1\,L_{\rm j}$, approximately as in FR~II
galaxies, and $R_{\rm blr}$ has been derived using Eq.~(\ref{Rblr}). 

\begin{figure}
\includegraphics[angle=270, width=0.5\textwidth]{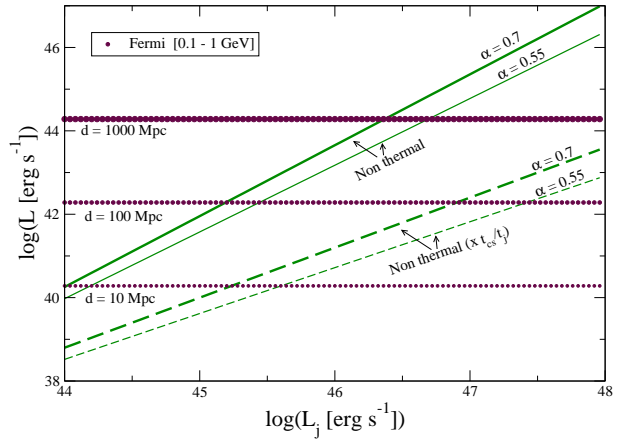}
\caption{Upper limits on the gamma-ray luminosity produced by 
$N_{\rm c}^{\rm j}$ clouds inside the jet as a function of
$L_{\rm j}$ in FR~II sources. Two cases are plotted, one assuming that 
clouds cross the jet without disruption (green solid lines), and one
in which the clouds are destroyed in a time as short as $t_{\rm cs}$
(green dashed lines). 
 The thick (solid and dashed) and thin (solid and dashed) lines 
correspond to 
$R_{\rm blr} \propto L_{\rm blr}^{0.7}$ (Kaspi et al. 2007) and 
$R_{\rm blr} \propto L_{\rm blr}^{0.55}$ (Kaspi et al. 2005), respectively.
In addition, the sensitivity levels of {\it Fermi} in
the range 0.1--1~GeV (maroon dotted lines) are plotted for three 
different distances $d=10$, 100 and 1000~Mpc.}
\label{L_tot}
\end{figure}

In Fig.~\ref{L_tot}, we show estimates for the gamma-ray luminosity
when many clouds interact simultanously with the jet. For this, we
have followed a simple approach assuming that most of the non-thermal
luminosity goes to gamma rays. This will be the case as long as the
escape and synchrotron cooling time are longer than the IC cooling
time (EC+SSC) for the highest electron energies. Given the little
information for the BLR in the case of FR~I sources, we do not
specifically consider these sources here. 

In the two next sections, we present more detailed calculations 
applying the model presented in Sect.~\ref{acc} to two
characteristic sources, Cen~A (FR~I, one interaction) and 3C~273 
(FR~II, many interactions).

\section{Application to FR~I galaxies: Cen~A} \label{FRI}

Cen~A is the closest AGN, with a distance $d\approx 3.7$~Mpc (Israel 1998). 
It has been classified as an FR~I radio galaxy and as a Seyfert~2 optical
object. The mass of the black hole is $\approx 6\times
10^7$~$M_{\odot}$ (Marconi et al. 2000). 
The angle between the jets and the line of sight is
large, $>50^\circ$ (Tingay et al. 1998), 
thus the jet radiation towards the observer  should
not suffer strong beaming. The jets of Cen~A are disrupted at kpc
scales, forming two giant radio lobes that extend $\sim 10^{\circ}$ in
the southern sky.  At optical
wavelenghts, the nuclear region of Cen~A is obscured by a dense region
of gas and dust, probably as a consequence of a recent merger (Thomson
1992,  Mirabel et al. 1999). At higher energies, {\it Chandra} and
{\it XMM-Newton} detected continuum X-ray emission coming from the
nuclear region with a luminosity $\sim 5\times10^{41}$~erg~s$^{-1}$
between 2--7~keV (Evans et al. 2004). These X-rays could be produced by the
accretion flow and the inner jet, although their origin is still
unclear. In the GeV range, Cen~A was detected above $200$~MeV by
\emph{Fermi}, with a bolometric luminosity of $\approx
4\times10^{40}$~erg~s$^{-1}$  (Abdo et al. 2009), and  above $\sim
200$~GeV by HESS, with a bolometric luminosity of $\approx
3\times10^{39}$~erg~s$^{-1}$ (Aharonian et al. 2009). In both cases,
this HE emission is associated with the nuclear region. 
Cen~A has been proposed to be a source of ultra HE cosmic rays (Romero
et al. 1996). 

A BLR has not been detected so far in Cen A (Alexander et al. 1999),
although this could be a consequence of the optical obscuration
produced by the dust lane. One can still assume that clouds surround
the SMBH in the nuclear region (Wang et al.  1986, Risaliti et
al. 2002) but, as a consequence of the low luminosities of the
accretion disc, it is not expected that the photoionization of these
clouds will be efficient enough to produce lines. Since no emission
from these clouds is assumed, we only consider the EC scattering with
photons from the accretion flow. 

\begin{figure}
\includegraphics[angle=270, width=0.5\textwidth]{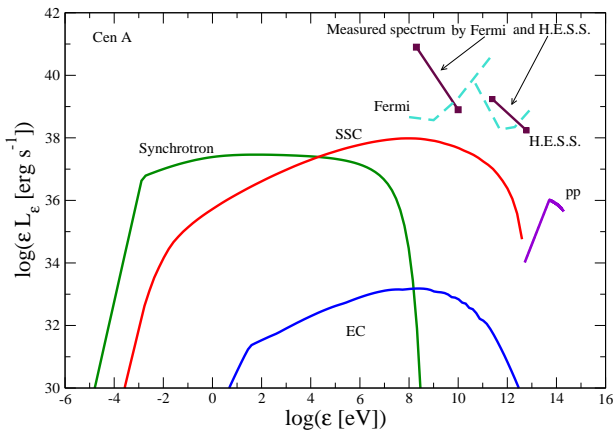}
\caption{Computed SED for one jet-cloud interaction at $z_{\rm int}$ in
Cen~A. We show also the SEDs of the detected emission by \emph{Fermi} and HESS, 
as well as the sensitivity curves of these instruments.}
\label{CenA}
\end{figure}

We adopt here a jet power for Cen~A of $L_{\rm j}=10^{44}$~erg~s$^{-1}$. 
From this value, and the values given
in Sect.~\ref{jet-cloud} for the remaining parameteres of the jet and
the cloud, $z_{\rm int}$ results in $\approx 5\times10^{15}$~cm. At
this jet height, the emission produced by the interaction between one
cloud and the jet is calculated assuming a $\eta_{\rm B}=0.01$, and
the corresponding SED is presented in Fig.~\ref{CenA}. As mentined in
Sect.~\ref{acc}, the low-energy band of the synchrotron
spectrum is self-absorbed at energies below $\sim 10^{-4}$~eV. At
gamma-ray energies, photon-photon absorption is negligible due to the
weak ambient photon fields (e.g. Rieger \& Aharonian 2009, Araudo et
al. 2009, 2010a,b). At high energies, SSC dominates the radiative output, with
the computed luminosity above 100~MeV being $\sim
2\times10^{39}$~erg~s$^{-1}$, and above 100~GeV about 10 times
less. These luminosities are below the sensitivity of \emph{Fermi} and
HESS and one order of magnitude smaller than the observed ones. Note
however that $L_{\rm nt}\propto R_{\rm c}^2$, and for slightly bigger
clouds, $L_{\rm nt}$ may grow up to detectable levels. The penetration
of a big clump in the base of the jet of Cen~A would lead to a flare
with a duration of about one day.

\section{Application to FR II galaxies: 3C~273 (off-axis)}\label{FRII}

\begin{table}[]
\begin{center}
\caption{Adopted parameters for Cen A and 3C~273.}
\label{applications}
\begin{tabular}{lcc}
\hline 
{}    & Cen A &  3C 273  \\ 
\hline 
Distance [Mpc] & 3.7 &  $6.7\times10^2$\\
Black hole mass  [$M_{\odot}$] & $6\times10^7$& $7\times10^9$ \\
Inclination angle [$^\circ$] & $> 50$  & $\sim 15$ \\
Jet luminosity [erg~s$^{-1}$] & $10^{44}$ & $4\times10^{47}$\\  
Disc luminosity [erg~s$^{-1}$]&$5\times10^{41}$& $2\times10^{46}$ \\
Disc photon energy$^{\star}$ [eV] & $\sim 5\times10^3$ & 54  \\
BLR luminosity $L_{\rm blr}$ [erg~s$^{-1}$]& - & $4\times10^{45}$  \\
\hline
$^{\star}$ of the thermal component.
\end{tabular}
\end{center}
\end{table}

3C~273 is a powerfull radio-loud AGN at a distance of $d=6.7\times
10^2$~Mpc (Courvoisier 1998) with a SMBH mass 
$M_{\rm BH} \sim 7\times10^9 M_{\odot}$
(Paltani \& T$\ddot{\rm u}$rler 2005). The angle of the jet with 
the line of sight is
small, $\approx 6^\circ$, which implies the blazar nature of 3C~273
(Jolley et al. 2009). The whole spectrum of this source shows 
variability (e.g. Pian et al. 1999) from years (radio) to few hours 
(gamma rays). 
At high energies, 3C~273
was the first blazar AGN detected in the MeV band by the COS-B
satellite and, later on, by EGRET (Hartman et
al. 1999). Recently, this source was also detected at GeV energies by
\emph{Fermi} and \emph{AGILE}, but it has not been detected yet in the
TeV range. Given the jet luminosity of 3C~273, $L_{\rm j}\approx
4\times10^{47}$~erg~s$^{-1}$ (Kataoka et al. 2002), $z_{\rm int}$ 
results in $\approx 3\times10^{17}$~cm. 
The BLR luminosity of this source is $\approx
4\times10^{45}$~erg~s$^{-1}$ (Cao \& Jiang 1999), and its size
$7\times10^{17}$~cm (Ghissellini et al. 2010), which implies that
jet-cloud interactions can take place. The disc luminosity is high,
$\approx 2\times 10^{46}$~erg~s$^{-1}$, with typical photon energies
$\approx 54$~eV (Grandi \& Palumbo 2004).

\begin{figure}
\includegraphics[angle=270, width=0.5\textwidth]{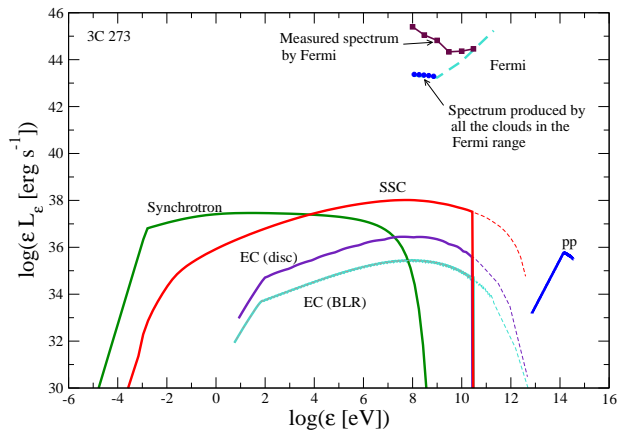}
\caption{Computed SED for one jet-cloud interaction at $z_{\rm int}$ in 3C~273. 
The emission in the 0.1--1~GeV range from many clouds inside the jet 
is also shown, together with the sensitivity level of \emph{Fermi} and 
the observed SED above 200~MeV.}
\label{3C273}
\end{figure}

The non-thermal SED of the radiation generated by jet-cloud
interactions in 3C~273 is shown in Fig.~\ref{3C273}. At $z_{\rm int}$,
the most important radiative processes are synchroton and SSC. The
bolometric luminosities by these processes in one interaction at
$z_{\rm int}$ are $6\times10^{38}$~erg~s$^{-1}$ and
$2\times10^{39}$~erg~s$^{-1}$, respectively. Given the presence of the
strong radiation fields from the disc and the BLR, the emission above
$\sim 10$~GeV is absorbed through photon-photon absorption,
and the maximum of the emission is around 0.1--1~GeV. Given
the estimated number of clouds in the BLR of 3C~273, $\sim 10^8$
(Dietrich et al. 1999), and the size is $R_{\rm blr}\approx
7\times10^{17}$~cm, the filling factor results in $f\sim 3\times
10^{-7}$. With this value of $f$, the number of clouds in the two jets
results in $\sim 2\times 10^3$ and $5\times 10^5$ for both the minimum
and the maximum values (see Sect.~\ref{Many-clouds}). Considering the
most optimistic case, the SSC luminosity would reach
$2\times10^{44}$~erg~s$^{-1}$.  This value is well below the observed
luminosity by \emph{Fermi} in the GeV range, $\sim
3\times10^{46}$~erg~s$^{-1}$ in the steady state and $\sim
1.7\times10^{47}$~erg~s$^{-1}$ in flare (Soldi et
al. 2009). However, the detected emission is very likely of beamed
origin and should mask any unbeamed radiation. However, powerful
non-blazar AGNs (FR~II galaxies) do not present this beamed component,
which makes possible the detection of GeV emission from jet-cloud
interactions in these sources. In this case, given that many BLR clouds 
can interact with the jet simultaneously, the radiation should be steady.

\section{Summary and discussion}\label{disc}

In this work, the interaction of clouds with the base of jets in AGNs
is studied. Considering reasonable cloud and jet parameters, we
estimate the relevant dynamical timescales of these interactions,
concluding that clouds can enter into the jet only above a certain
height, $\sim z_{\rm int}$. Below $z_{\rm int}$, the jet is too
compact and its ram/magnetic pressure will destroy the cloud before
fully penetrating into the jet. Once the cloud significantly interacts
with the jet, strong shocks are generated and gamma rays can be
produced with an efficiency that depends strongly on the bow-shock
magnetic field. 

Bow shock $B$-values well below equipartition with non-thermal
particles allow significant gamma-ray emission. For very high
$B$-values (Poynting-flux dominated jets), the treatment performed
here does not apply. In that case, $z_{\rm int}$ could be still
defined adopting the jet magnetic pressure instead of the ram
pressue. If a cloud entered in such a jet, particle acceleration in
the bow shock could still occur due to, for instance, magnetic
reconnection. The study of such a case would require a completely
different approach than the one presented here. In general, for
bow-shock magnetic fields above equipartition with
non-thermal particles, the IC channel and gamma ray production will be
supressed in favor of synchrotron emission. Unless magnetic
dissipation reduced the magnetic field enough for IC to be
dominant.  Bow shock $B$-values well below equipartition with
non-thermal particles allow significant gamma-ray emission. We note
that modeling of gamma rays from AGN jets uses to require relatively
low magnetic fields (e.g. Ghisellini et al. 2010). Therefore, it could
be that, even if the jet magnetic field were high at $z_{\rm int}$, it
could become small enough farther up due to bulk acceleration
(e.g. Komissarov et al. 2007) or some other form of magnetic
dissipation.

For very nearby sources, like Cen~A, the interaction of big clouds
with jets may be detectable as a flaring event, although the number of
these big clouds and thereby the duty cicle of the flares are difficult
to estimate. Given the weak external photon fields in these sources,
VHE photons can escape without suffering significant
absorption. Therefore, jet-cloud interactions in nearby FR~I may be
detectable in both the HE and the VHE range as flares with timescales
of about one day.  Studying such a radiation would provide information
on the environmental conditions and the base of the jet in these
sources.

In FR~II sources, many BLR clouds could interact simultaneously with
the jet. The number of clouds depends strongly on the cloud lifetime
inside the jet, which could be of the order of several $t_{\rm cs}$. 
Nevertheless, it is worthy noting that after cloud fragmentation
many bow shocks can still form and efficiently accelerate particles if
these fragments are slower than the jet. Since FR~II sources are
expected to present high accretion rates, radiation above 1~GeV
produced in the jet base can be strongly attenuated due to the dense
disc and the BLR photon fields, although gamma rays below 1~GeV should
not be significantly affected. Since jet-cloud emission should be
rather isotropic, it would be masked by jet beamed emission in blazar
sources, although since powerful/nearby FR~II jets do not present
significant beaming, these objects could indeed show gamma rays from
jet-cloud interactions. In the context of AGN unification (Urry \&
Padovani 1995), the number
of non-blazar (radio-loud) AGNs should be much larger than that of
blazars with the same $L_{\rm j}$. As shown in Fig.~\ref{L_tot}, close
and powerful sources could be detectable by deep enough observations
of {\it Fermi}. After few-years exposure a significant signal from these
objects could arise, their detection providing strong evidence that
jets are already strongly matter dominated at the bow shock regions, as well
as physical information on the BLR and jet base region.  

\begin{acknowledgements}
We thank the referee Elena Pian for constructive comments and sugestions.  
A.T.A. and V.B-R. thanks Max Planck Institut fuer Kernphysik for kind 
hospitality and support. A.T.A. and G.E.R. are supported by CONICET 
and the Argentine agency ANPCyT (grant BID 1728/OC-AR PICT 2007-00848).
V.B-R. and G.E.R. acknowledge support by the Ministerio de 
Educaci\'on y Ciencia (Spain) under grant AYA 2007-68034-C03-01, FEDER funds. 

\end{acknowledgements}

{}
\end{document}